\documentclass[aps,pra,twocolumn]{revtex4-2}

\pdfoutput=1

\usepackage{amsfonts,amsmath,amssymb}

\usepackage{graphicx}
\graphicspath{{graphics/}}

\usepackage{lmodern} 
\usepackage[T1]{fontenc}
\usepackage[utf8]{inputenc} 
\usepackage{nicefrac}
\usepackage{textcomp}
\usepackage{braket}
\usepackage{slashed}
\usepackage{tensor}
\usepackage[subnum]{cases}
\usepackage{mathtools}
\usepackage{dsfont}

\usepackage{xcolor}

\newcommand{\de}{\partial}
\newcommand{\del}{\nabla}

\newcommand{\h}{\hbar}

\renewcommand{\a}{\alpha}
\renewcommand{\b}{\beta}

\renewcommand{\b}{\beta}

\newcommand{\T}{\Theta}
\newcommand{\w}{\omega}

\newcommand{\vc}[1]{\mathbf{#1}}

\renewcommand{\L}{\mathcal{L}}

\newcommand{\A}{\mathcal{A}}
\newcommand{\B}{\mathcal{B}}

\newcommand{\abs}[1]{\left| #1 \right|}

\newcommand{\tld}[1]{\widetilde{#1}}

\begin{document}

\title{Spontaneous quantum superradiant emission in atomic Bose--Einstein condensates subject to a synthetic vector potential}

\author{Luca Giacomelli}
\email[]{luca.giacomelli-1@unitn.it}
\author{Iacopo Carusotto}
\email[]{iacopo.carusotto@unitn.it}
\affiliation{INO-CNR BEC Center and Dipartimento di Fisica, Universit\`a di Trento, via Sommarive 14, I-38050 Povo, Trento, Italy}

\begin{abstract}
	We theoretically investigate the spontaneous quantum emission of phonon pairs by superradiant processes in an atomic Bose--Einstein condensate subject to a synthetic vector potential. Within the analog gravity perspective, this effect corresponds to the spontaneous emission of radiation from the ergosurface of rotating black holes. A general input-output formalism is built and used to characterize the spectral and correlation properties of the emission. Experimentally accessible signatures of the emission are pointed out in the correlation functions of the atomic gas.
\end{abstract}

\maketitle

\section{Introduction}

Rotating black holes in general relativity are known to allow superradiant scattering, namely the amplified reflection of radiation that impinges on them~\cite{brito2020superradiance}. This amplification can be seen as a stimulated emission of radiation in response to the incident field. The same works making the very first qualitative prediction of superradiant scattering~\cite{zeldovich1,zeldovich2} also pointed out how at the quantum level the stimulated emission must be associated to a \textit{spontaneous} emission in the superradiant modes. This anticipation was then formally confirmed using a second quantized formalism  for bosonic fields on the Kerr spacetime~\cite{unruh1974second,ford1975quantization}. As a first example of quantum spontaneous emission from black holes, it served as an inspiration for the discovery of Hawking radiation \cite{hawking1975particle}. A historical perspective on these fascinating developments can be found in ~\cite{page2005hawking}.

While of high importance for theoretical understanding of gravity, the weakness of the quantum emission from astrophysical objects  makes its experimental observation  extremely challenging. A different route to investigate the physics behind these phenomena was proposed by the analog gravity programme~\cite{barcelo2011analogue}, which aims at realizing condensed matter systems that reproduce the physics of curved spacetimes in realizable and controllable laboratory setups.

This idea was first proposed for the study of analog Hawking radiation~\cite{unruh1981experimental}, namely the spontaneous emission of sound waves from an \textit{acoustic horizon} separating regions of sub- and super-sonic flow in a non-uniformly flowing fluid. 
Such an analog Hawking emission was recently observed in a Bose--Einstein condensate (BEC) of ultracold atoms displaying such a flow configuration~\cite{steinhauer2016observation,de2019observation}. This amazing success was made possible by the extremely low temperature of the sample, at which quantum features of sound become visible~\cite{finazzi2014entangled}, and by the possibility of performing measurements also in the interior of the black hole: The key signature of Hawking emission consisted in fact of the correlations between density fluctuations on opposite sides of the horizon as theoretically anticipated in~\cite{balbinot2008nonlocal,carusotto2008numerical}.

The same strategy can be applied to superradiance effects. The direct analog of a rotating black hole is the so called \textit{draining bathtub vortex}~\cite{barcelo2011analogue}, combining a radial and an azymuthal flow: Besides the acoustic horizon, this kind of flow displays an \textit{acoustic ergoregion} (delimited by the so-called \textit{ergosurface}), that supports the negative-energy modes at the basis of amplified reflection. Such a configuration was recently employed in the first observation of classical superradiant scattering using surface gravity waves on water~\cite{torres2017rotational}. Of course, the high temperature of the system hinders any exploration of quantum effects. Also from the theoretical point of view, quantum features of superradiance in analogs have received a limited attention: Superradiance-induced friction effects  on a rotating body immersed in a superfluid were anticipated in~\cite{calogeracos1999rotational}. Recently a second-quantized procedure was applied to quantum superradiance for a dispersive field in a draining bathtub configuration~\cite{patrick2020rotational}.

In spite of their remarkable success in pioneering investigations of analog superradiant scattering, vortex-like rotating configurations have a limited tunability in the flow parameters that can be realistically obtained. Moreover, the cylindrical geometry restricts the available space making it difficult to detect quantum emission from correlation functions. To solve these issues, in the recent work~\cite{giacomelli2021understanding}, we proposed an alternative kind of analog system where to investigate superradiance. Instead of a rotating fluid, we considered a local tuning of the velocity by means of a synthetic vector potential \cite{dalibard2011colloquium}: This trick removes the usual irrotationality constraint of superfluids and allows to realize arbitrary  \textit{rotational} flows. As a most promising configuration, we considered a \textit{planar} ergosurface separating two regions of uniform flow parallel to the interface but different velocities. Removing the irrotationality constraint allows to decouple the different elements at play in superradiant scattering, in particular to realize ergoregions without horizons and without dynamical instabilities. In this way, we could develop an intuitive picture of superradiant scattering and of the associated instabilities at the classical level~\cite{giacomelli2021understanding}.

In the present work, we make a further step towards the development of a general theory of quantum superradiant effects in such systems. In particular, we extend the input-output theory of spontaneous quantum emission originally introduced in~\cite{recati2009bogoliubov} for Hawking radiation from analog black hole horizons in one-dimensional flows to higher-dimensional geometries. 
As a first application of our theory, we investigate the quantum emission in the geometry with two regions of uniform and parallel flows separated by a step-like planar interface. In its simplicity, this configuration provides a useful toy model of an ergosurface which correctly accounts for classical superradiant scattering, as we discussed in \cite{giacomelli2021understanding}, and for quantum pair production at the ergosurface, as we are going to show in this work.
Most importantly, the geometric simplicity of our setup allows to characterize the quantum emission in terms of correlations between fluctuations on the two sides of the ergosurface. Like for analog Hawking radiation, this turns out to be a powerful tool to experimentally detect the spontaneous production of phonon pairs by spontaneous superradiance processes. In contrast to the celebrated moustache-shaped feature of Hawking radiation, no dramatic feature appears in the position-space density-density correlations but a non-trivial pattern is predicted for the momentum-space two-particle correlations.

The structure of the article is the following. In Section \ref{sec:system} we introduce the physical system under investigation and we describe how an acoustic spacetime displaying a half-plane ergoregion can be obtained by means of a suitably designed synthetic vector potential. In Section \ref{sec:dispersions} we review the Bogoliubov dispersion of the excitations and we summarize the kinematics of superradiant scattering in our configuration. The reader that is familiar with our previous work~\cite{giacomelli2021understanding} can quickly fly through these first sections and focus on the novel results that are discussed in the following Sections. 
In Section \ref{sec:scattering-solutions} we construct the scattering modes and we develop the high-dimensional generalization of the input-output formalism for the quantum emission. In Section \ref{sec:scattering-matching} we numerically evaluate the scattering coefficients for the simplest case of a step-like transition at the ergosurface: This gives a further characterization of the different regimes of classical superradiant scattering and provides a quantitative prediction for its amplitude. In Section \ref{sec:quantization}, we make use of the input-output theory at the quantum level: We identify the different channels for spontaneous pair production into the superradiant modes and we characterize the spectrum of the emission. In Section \ref{sec:corr} we investigate the signatures of the spontaneous superradiant emission in  the correlation functions and we propose strategies for experimental observation: While the features in the density-density correlations are too complex for a straightforward analysis, two-body correlation function in momentum space are predicted to display non-trivial yet transparent features. Conclusions are finally drawn in Sec.\ref{sec:conclu}.


\section{The physical system: A planar ergosurface}
\label{sec:system}
As in our previous work~\cite{giacomelli2021understanding}, we consider an atomic BEC tightly confined in one direction, so that one dimension is frozen and the relevant dynamics takes place in two spatial dimensions only. For weak enough interactions to be in the dilute regime and assuming a vanishing initial temperature, the condensate can be described at the mean-field level in terms of a complex scalar scalar field $\Psi(x,y,t)$ obeying the two-dimensional Gross--Pitaevskii equation (GPE) \cite{pitaevskii2016bose}, 
\begin{equation}\label{eq:gpe_gauge}
	i\h \de_t \Psi = \left[\frac{(-i\h\del-\vc{A})^2}{2M}+V+g\abs{\Psi}^2-\mu\right]\Psi\,.
\end{equation}
were $M$ is the atomic mass, $V$ is an external trapping potential, $g$ is the atom-atom interaction constant, and $\mu$ is the chemical potential. Furthermore, the condensate is assumed to be coupled to a (synthetic) vector potential $\mathbf{A}$. As it is reviewed in~\cite{spielman2009raman}, vector potentials with a variety of different spatio-temporal shapes can be  obtained with suitable combinations of optical and/or microwave and/or magnetostatic fields that have the effect of shifting the position of the minimum of the effective dispersion relation of the atoms. As a consequence, the physical velocity
\begin{equation}
	\vc{v}=\frac{\h\,\del\T - \vc{A}}{M}=\vc{v}_{\rm can}-\frac{\vc{A}}{M},
\end{equation}
differs from the canonical velocity $\vc{v}_{\rm can}$ given by the gradient of the condensate phase and, in particular, is no longer constrained to be irrotational~\cite{leblanc_spielman_2017}, as it instead happens in usual superfluid hydrodynamics. Since analog gravity is based on a geometric description of sound propagation in a moving fluid in terms of a curved-spacetime metric and this latter depends on the velocity field~\cite{barcelo2011analogue}, the possibility of having a rotational flow greatly expands the set of spacetimes that analog models can realize.

In this article we focus our attention on the simplest configuration introduced in~\cite{giacomelli2021understanding} providing a minimal \textit{toy model} displaying superradiant scattering. Both the synthetic vector field $\vc{A}$ and the external potential $V$ are taken to only depend on the $y$ coordinate, so that the system is translationally invariant along $x$. In particular, we take the vector potential $\vc{A}(y)$ to be everywhere directed along $x$ and we include an external potential $V(y)=-A_x^2(y)/2M$, so that a condensate in the plane-wave form
\begin{equation}\label{eq:stationry-state}
	\Psi_0(x,y)=\sqrt{n}e^{iKx}
\end{equation}
is a stationary state of the GPE. This state describes a BEC of constant density $n$ flowing along $x$ with a $y$-dependent (and thus rotational) velocity field $v_x(y)=[\hbar K - A_x(y)]/M$.  The speed of sound is instead constant and equal to $c_s=\sqrt{gn/M}$. 

From the point of view of the acoustic spacetime, any region of supersonic motion is an ergoregion \cite{barcelo2011analogue} and any interface separating a subsonic region from a supersonic one is an ergosurface. An acoustic horizon corresponds instead to an interface  delimiting a region in which the component of the velocity normal to the interface is supersonic, so that sound cannot exit that region. This region then corresponds to the interior of the acoustic black hole. In this work, we focus on configurations in which the spatial variation of the velocity only occurs in the direction perpendicular to the velocity, which excludes the presence of acoustic horizons. In particular, we consider configurations in which the physical velocity approaches constant values far away from the $y=0$ line and increases from a \textit{slow} (s) subsonic value $v_x^s<c_s$ for $y<0$ to a \textit{fast} (f) supersonic value $v_x^f>c_s$. This describes an acoustic spacetime with a \textit{planar} ergosurface at $y=0$ and no horizon.

Ergoregions in rotating black holes are known to support field excitations with a negative energy with respect to an asymptotic observer. Such excitations are the basic ingredient for superradiant scattering: A positive-energy wave coming from infinity can in fact be transmitted into the ergoregion as a negative-energy wave, with the consequence that, by energy conservation, the reflected positive-energy wave will have a larger amplitude than the incident wave. Since our setup is invariant for Galilean boosts along $x$, if the velocity difference between the two regions is such that $\Delta v_x:=v_x^f-v_x^s<2c_s$, a reference frame can be found where the flow is everywhere subsonic and no negative-energy excitation is present. As a result, superradiant scattering in this configuration is only possible for $\Delta v_x>2c_s$.

\begin{figure}[t]
	\centering
	\includegraphics[width=\columnwidth]{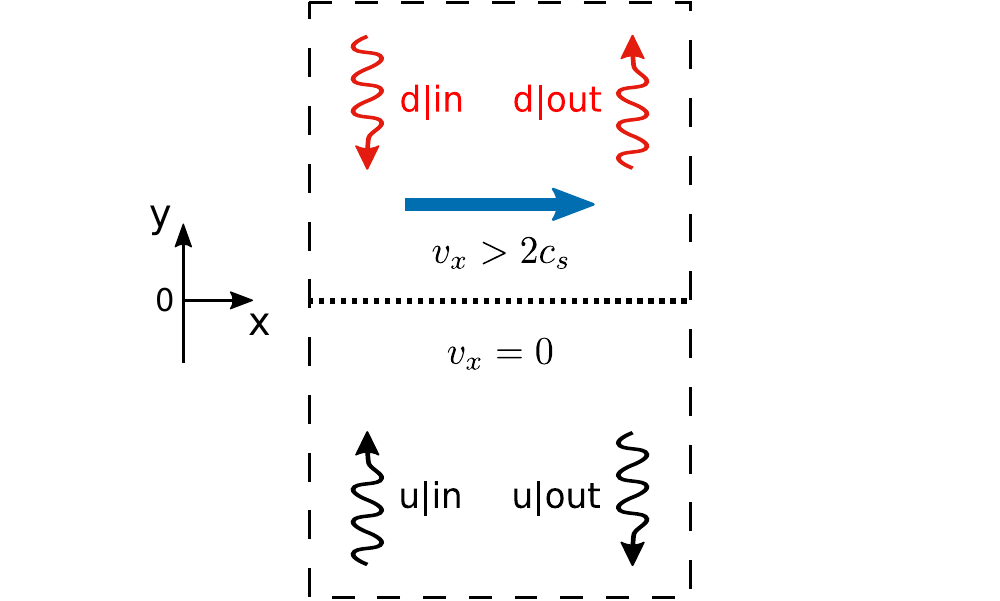}
	\caption{Sketch of the physical configuration under examination. The condensate velocity is induced by the synthetic vector potential only in the upper $y>0$ part of the system, while the lower $y<0$ part is at rest. The wiggly lines show the propagation direction of the modes indicated by dots on the dispersion curves of Figure \ref{fig:superradiant-dispersion} and involved in the superradiant processes.
	}
	\label{fig:diagram}
\end{figure}

For simplicity, in what follows we restrict our attention to the $K=0$ case in which the condensate wavefunction $\Psi_0$ can be taken as a real constant: The condensate velocity is then fixed by the external synthetic vector field. In particular, we assume that the asymptotic slow region is at rest $v_x^s=-A_x^s/M=0$, while the fast one has a speed $v_x^f=-A_x^f/M$. This configuration is sketched in Figure \ref{fig:diagram}. 

Before proceeding, some comments on the stability of the configuration under investigation are in order. From general relativity, it is known that spherically symmetric spacetimes displaying an ergoregion and no horizon are subject to ergoregion instabilities. On the other hand, such instabilities get quenched in the presence of a horizon: The horizon plays the role of an open boundary condition for field fluctuations, preventing the localization of the negative-energy waves in the ergoregion and suppressing the possibility of a repeated amplification. A similar mechanism underlies the stability of the configuration under investigation here: The system is assumed to be infinite along the $y$ direction, so that open boundary conditions are automatically enforced on both sides of the ergoregion. This guarantees an efficient evacuation of the field fluctuations, so that no ergoregion instabilities can take place~\cite{giacomelli2021understanding}.


\section{The kinematics of superradiant scattering}
\label{sec:dispersions}

In this Section, we briefly review some basic concepts on superradiant phenomena in the specific configuration under examination here. The reader that is already familiar with our previous work~\cite{giacomelli2021understanding} can quickly skim this Section and focus the attention on the following ones where our new results are presented.

As usual in analog models based on Bose--Einstein condensates, the quantum field living in the curved spacetime physically corresponds to small amplitude fluctuations on top of the BEC. These are described with the so-called Bogoliubov theory~\cite{pitaevskii2016bose}, in which one considers a small perturbation $\Psi_0+\delta\psi$ of the GPE stationary state \eqref{eq:stationry-state} and one focuses on a linearized form of the GPE dynamics \eqref{eq:gpe_gauge}. The resulting linear problem can be described in terms of the spinor $(\delta\psi,\,\delta\psi^*)^T$, in which the fluctuation field $\delta\psi$ and its complex conjugate $\delta\psi^*$ are taken as independent variables~\cite{castin2001bose}.

Sufficiently far from the ergosurface, our setup is composed by two regions where the condensate is uniform. Within each of them, the linear perturbations can be decomposed in eigenstates of the momentum $\vc{k}$, 
\begin{equation}
	\begin{pmatrix}
		\delta\psi \\ \delta\psi^*
	\end{pmatrix}
	(x,y)
	= e^{ik_xx+ik_yy}
	\begin{pmatrix}
		U_\vc{k} \\ V_\vc{k}
	\end{pmatrix}
	:=e^{ik_xx+ik_yy}\ket{\vc{k}}.
\end{equation}
Within each region, the linear Bogoliubov eigenvalue problem has the form $\h\w_\vc{k}^{s,f}\ket{\vc{k}}=\L^{s,f}\ket{\vc{k}}$, with
\begin{equation}\label{eq:bogo-problem}
	\L^{s,f}=
	\begin{bmatrix}
		D_{\vc{k}} +\h v_x^{s,f}k_x & gn \\
		\\
		-gn & -D_{\vc{k}} +\h v_x^{s,f}k_x
	\end{bmatrix}
\end{equation}
and $D_{\vc{k}}:=\frac{\h^2\vc{k}^2}{2M} + gn$.

The eigenvalues of this matrix give the dispersion relation of fluctuations in each region,
\begin{equation}\label{eq:dispersion}
	\h\w_\vc{k}^{s,f} = \h v_x^{s,f}k_x \pm \B(k_x,k_y),
\end{equation}
where  the first term describes the Doppler shift of the excitations by the moving condensate and
\begin{equation}\label{eq:bogo-term}
	\B(\vc{k}):= \sqrt{\frac{\h^2\vc{k}^2}{2M}\left(\frac{\h^2\vc{k}^2}{2M} + 2gn\right)}
\end{equation}
is the celebrated Bogoliubov dispersion relation for a uniform condensate at rest~\cite{pitaevskii2016bose}. The plus and minus signs in \eqref{eq:dispersion} refer to the two positive and negative norm eigenmodes of \eqref{eq:bogo-problem}, where the Bogoliubov norm is given by $|U_\vc{k}|^2-|V_\vc{k}|^2$. The energy of an eigenmode is 
given by $E_\vc{k}=\h\w_\vc{k}(|U_\vc{k}|^2-|V_\vc{k}|^2)$, so that negative-norm modes at positive frequencies have a negative energy.

For momenta $|\vc{k}|\ll \sqrt{Mgn}/\h=:1/\xi$, with $\xi$ the so-called healing length of the condensate, the Bogoliubov dispersion relation \eqref{eq:bogo-term} reduces to a sonic dispersion $\B(\vc{k})\simeq \h c_s\,|\vc{k}|$. For higher momenta $|\vc{k}|\gg 1/\xi$ the Bogoliubov eigenmodes have a group velocity larger than the speed of sound and, in analogy with the speed of light, the dispersion relation is called \textit{superluminal}. The limit in which the Bogoliubov dispersion relation can be accurately approximated with the sonic dispersion is called the \textit{hydrodynamic limit}. Here, the Bogoliubov linear problem reduces to a Klein--Gordon equation in a curved spacetime. Even though this is the only regime in which strictly speaking the gravitational analogy holds, in this work we will consider superradiance in the fully dispersive case and we will comment on the differences with the Klein--Gordon case when needed.

Since our system is translationally invariant along $x$, $k_x$ is a conserved quantity and we can treat the different $k_x$ channels as separate one-dimensional problems along $y$. Note that the Bogoliubov problem of the whole system has a \textit{particle-hole symmetry} expressed by the fact that the spectrum for $-k_x$ is specular to the one for $k_x$; that is there are pairs $i,j$ of eigenmodes at $\pm k_x$ that are related by
\begin{equation}\label{eq:particle-hole}
    \begin{pmatrix}
        U_{-k_x,j}\\ V_{-k_x,j}
    \end{pmatrix}
    =
    \begin{pmatrix}
        V_{k_x,i}\\ U_{k_x,i}
    \end{pmatrix};
    \hspace{0.5cm}
    \w_{-k_x,j}=-\w_{k_x,i}.
\end{equation}
Based on this symmetry, we can restrict our following analysis to positive values of $k_x$. 

\begin{figure}[t]
	\centering
	\includegraphics[width=\columnwidth]{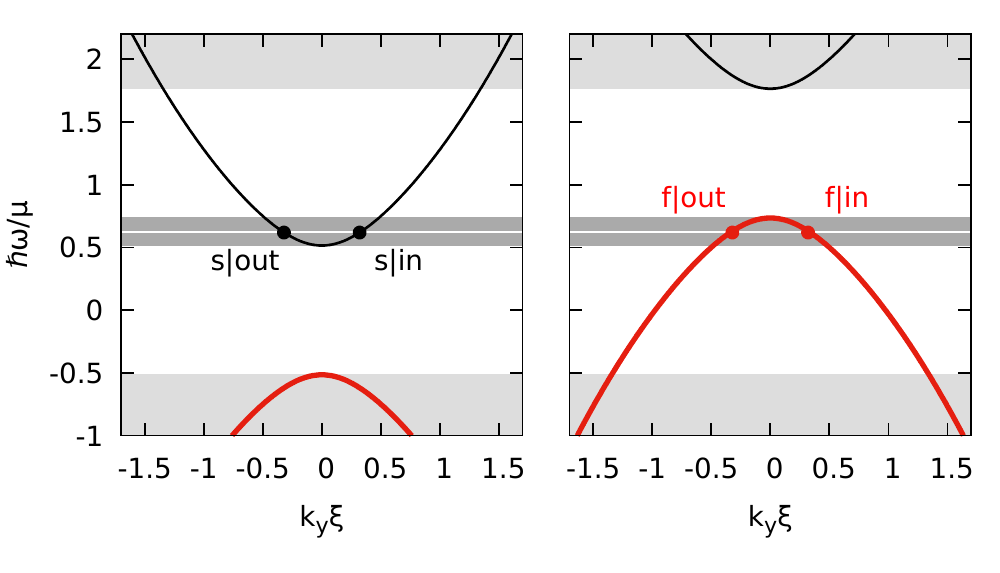}
	\caption{Dispersion relations of the Bogoliubov modes at a given $k_x\xi=0.5$ in the two regions. The left and right panels correspond to the slow and fast regions, respectively with $v_x^s=0$ and $v_x^f=2.5\ c_s$. Black thin and red thick lines correspond to positive and negative norm modes. The light gray regions indicate the frequency ranges where ordinary scattering occurs, while the dark grey ones indicate the superradiant scattering range. The white line indicates an example of frequency in the superradiant range; the dots on the dispersion curves indicate the modes involved in the superradiant scattering process.}
	\label{fig:superradiant-dispersion}
\end{figure}

Examples of cuts of the dispersion relations $\w_{\vc{k}}^{s,f}$ at fixed $k_x$ in the two regions are shown in Figure \ref{fig:superradiant-dispersion}. The left plot refers to the $y<0$ region where the condensate is not moving: Here, one can see that the effect of a transverse momentum is to introduce a gap in the dispersion relation, so that we can think of our field as having a mass. The right plot refers instead to the fast region at $y>0$: Here, the effect of the Doppler term in \eqref{eq:dispersion} is to vertically shift the dispersion relation, so that some negative-norm modes are pushed to positive frequencies and thus acquire negative energies.

Within  each uniform region the fluctuation field can be written as a linear combination of plane-wave eigenmodes. Since the Bogoliubov dispersion relation \eqref{eq:dispersion} is of fourth order in the momenta $k_y$, it will generally have 4 roots for fixed values of the conserved frequency $\w$ and $k_x$ momentum component. For the real frequencies considered here, these roots can either be divided in a pair of real roots and a pair of complex conjugates ones or are all complex forming two pairs of complex conjugate roots. In the configurations considered in this work, it never happens to have four real roots. This could occur if a supersonic $y$ component of the velocity was also present.

Since we are considering an infinite system along the $y$ direction, we need to impose that the modes are bounded at infinity, which implies that the imaginary part of $k_y$  must be non-negative in the upper region and non-positive in the lower region. As a result, if no root is real, there will be two relevant (evanescent) modes; on the other hand, if two roots are real, there will be one evanescent mode and two more propagating modes, for a total of three physically relevant modes. In general, 
as one can see in Figure \ref{fig:superradiant-dispersion}, there are no real roots for frequencies $\w$ in the \textit{mass gap}
\begin{equation}\label{eq:massgap}
	\h v_x^{s,f}k_x -\B(k_x)< \h\w < \h v_x^{s,f}k_x + \B(k_x),
\end{equation}
while two real roots are present for $\w$ above (below) the gap, corresponding to positive (negative) norm modes.

In our study of superradiant physics, we are mostly interested in scattering configurations where a plane wave comes from infinity in one of the two regions and is incident on the transition region around $y=0$, getting then reflected and transmitted into other modes. The different kinds of scattering that can occur can be graphically identified from Figure \ref{fig:superradiant-dispersion}, simply by looking at the available modes at each frequency. In analogy to superradiant scattering in black holes, where the ingoing wave has a positive energy and comes from outside the ergoregion, we focus here on the case of positive-norm ingoing waves incident from the slow region. As mentioned in~\cite{giacomelli2021understanding}, the situation in which the ingoing wave comes from the fast region would however be completely analogous.

For frequencies above $\h v^f_xk_x +\B(k_x)$, positive-norm modes are available in both regions and the incoming wave will be partially transmitted and partially reflected. An analogous behaviour occurs for $\h\w<-\B(k_x)$ where negative-norm modes are available in both regions. In~\cite{giacomelli2021understanding}, we called this kind of same-norm scattering \textit{ordinary scattering}, since no amplification can occur. These regions are indicated in Figure \ref{fig:superradiant-dispersion} with a light gray shading.
For frequencies in the mass gap of the fast region there is no propagating mode available for transmission, so the incident wave coming from the slow region will only couple to the evanescent modes and will thus be \textit{totally reflected}.

The most interesting regime occurs for
\begin{equation}
	\B(k_x)<\h\w<\h v^f_xk_x -\B(k_x)\,:
\end{equation}
In this range, the incoming positive-norm wave can be transmitted into the fast region as a negative energy wave. Energy conservation then implies a corresponding amplification of the reflected wave. The frequency interval in which this \textit{superradiant scattering} occurs is indicated in Figure \ref{fig:superradiant-dispersion} with a dark gray shading. 

Hence, superradiant scattering is only possible if (for positive $k_x$) the maximum of the lower branch in the fast region is at a higher frequency than the minimum of the upper branch in the slow region; this requires that
\begin{equation}\label{eq:superradiant-condition}
	\h v_x^f k_x> 2\B(k_x)\,.
\end{equation}
In the hydrodynamic limit, this reduces to the $v_x^f> 2c_s$ condition mentioned above. For the full superluminal Bogoliubov dispersion, this condition depends instead on $k_x$: This introduces an upper bound on the transverse momentum, above which no superradiant scattering is possible
\begin{equation}\label{eq:dispersive-threshold}
	k_x<\frac{1}{\xi}\sqrt{\frac{(v_x^f)^2}{c_s^2}-4}=:k_x^{max}.
\end{equation}
As expected, the explicit dependence on $\xi$ of this expression confirms that the upper bound is not present in the non-dispersive case.

\section{The structure of the scattering solutions}
\label{sec:scattering-solutions}

As it was done in \cite{leonhardt2003bogoliubov,recati2009bogoliubov} for the case of analog Hawking radiation in one-dimensional condensates, a quantum theory of the superradiant scattering problem can be developed in terms of a scattering matrix connecting the operator-valued amplitudes of ingoing and outgoing modes. As compared to the Hawking case considered in the quoted works, our development here will include the lateral degrees of freedom: Assuming translational invariance along $x$, the transverse dynamics will be encapsulated in the conserved wavevector $k_x$.

As indicated in Figures \ref{fig:diagram} and  \ref{fig:superradiant-dispersion}, for each value of $k_x$ and of the frequency $\w$, we can distinguish in each region $I=s,f$ among the non-evanescent modes the \textit{ingoing} ones ($I|in$), whose $y$-component of the group velocity $v_{I|in}:=\left.\de_{k_y}\w_\vc{k}^I\right|_{k_y=k_{I|in}}$ is directed towards the interface, and \textit{outgoing} ones ($I|out$) whose (analogously defined) group velocity is directed away from it. 
Within each region, we can write the fluctuation field at frequencies at which there are real roots as 
\begin{equation}\label{eq:inout-modes}
	\begin{split}
	\begin{pmatrix}
		U(y)\\ V(y)
	\end{pmatrix}_{k_x,\w,I}
	=&
	\A_{I|in}
	\begin{pmatrix}
		U_{k_{I|in}}\\ V_{k_{I|in}}
	\end{pmatrix}
	\frac{e^{ik_{I|in}y}}{\sqrt{2\pi|v_{I|in}|}}\\
	+&
	\A_{I|out}
	\begin{pmatrix}
		U_{k_{I|out}}\\ V_{k_{I|out}}
	\end{pmatrix}
	\frac{e^{ik_{I|out}y}}{\sqrt{2\pi|v_{I|out}|}}\\
	+&
	\A_{I|ev}
	\begin{pmatrix}
		U_{k_{I|ev}}\\ V_{k_{I|ev}}
	\end{pmatrix}
	\frac{e^{ik_{I|ev}y}}{\sqrt{2\pi}},
	\end{split}
\end{equation}
where $I=s,f$ and the spinors on the right-hand side are normalized to $|U_k|^2-|V_k|^2=\pm 1$. A spatio-temporal plane-wave dependence as $e^{i(k_x x -\omega t)}$ is implicitly assumed for all fields. The chosen form of the normalization factors of non-evanescent modes involves the $y$-component of the group velocities $v_{I|in, out}$ guarantees that the full mode wavefunction at fixed frequency $\w$ is normalized to $\delta(\w)$. If no propagating modes are present at the frequency $\w$, one is left with the sum of two evanescent waves.

If not all modes are evanescent, a scattering solution is obtained by selecting one ingoing mode ($s|in$ or $f|in$) and setting all other ingoing amplitudes (if any) to zero. If propagating modes are present on both sides there will be a linear input-output relation connecting the outgoing modes amplitudes to the ingoing ones~\cite{leonhardt2003bogoliubov,recati2009bogoliubov}
\begin{equation}\label{eq:inputoutput}
\begin{pmatrix}
	\A_{s|out}\\ \A_{f|out}
\end{pmatrix}
=S(k_x, \omega)
\begin{pmatrix}
	\A_{s|in}\\ \A_{f|in}
\end{pmatrix}.
\end{equation}
The square moduli of the scattering matrix elements $|S_{IJ}(k_x,\w)|^2$ give the reflection or transmission amplitudes into the mode $I|out$ for an incoming wave in mode $J|in$. The chosen mode normalization guarantees that for each value of the frequency $\w$ and the transverse wavevector $k_x$, the scattering matrix $S(k_x,\w)$ satisfies the pseudo-unitarity property
\begin{equation}
	S^\dag \eta S = \eta,
\end{equation}
where $\eta$ is a diagonal matrix having as elements $+1$ for each positive-norm mode and $-1$ for each negative-norm one. This condition expresses the conservation of the Bogoliubov norm during the scattering and, thus, of the energy.

Focusing on an ingoing positive-norm mode from the slow region, $\eta=\sigma_3$ or $\eta=\mathds{1}$ depending on the sign of the norm of the modes in the fast region. In the former case, one readily sees the occurrence of superradiance: The conservation of the norm in the form
\begin{equation}
	|S_{ss}|^2-|S_{fs}|^2=1
\end{equation}
allows for a reflection coefficient $|S_{ss}|^2$ greater than one, provided it is compensated by a non-vanishing transmission $|S_{fs}|^2$: In this case, superradiance is directly visible as the intensity of the reflected wave exceeds the one of the incoming wave.

\section{Matching solution for a step-like ergosurface}
\label{sec:scattering-matching}

The approach described in the previous Section to build the scattering matrix can be applied to every configuration in which the system is translationally invariant along $x$ and the condensate flow speed approaches constant asymptotic values sufficiently far from the transition region. In analogy with the step configuration introduced in~\cite{recati2009bogoliubov} to describe analog Hawking radiation, a particularly simple situation is the one in which there is a step-like transition between two values of the transverse velocity $v_x(y)=v_x^f\Theta(y)$ (i.e. of the synthetic vector potential $A_x(y)=A_x^f\Theta(y)$), where $\Theta(y)$ is the Heaviside step function. In the following of the work we will focus on this configuration.

In the Hawking case, one may be concerned that a step-like transition corresponds to an formally infinite Hawking temperature for which the gravitational analogy is no longer valid. In spite of this, it was shown in~\cite{recati2009bogoliubov} that this simple model still provides an accurate description of the Hawking radiation process. This difficulty is absent in the present case of superradiant scattering, for which our earlier work~\cite{giacomelli2021understanding} has shown that a step-like transition does not have any dramatic effect. A smoother transition only results in a quantitative reduction of the superradiant scattering coefficients: Since the mode-conversion from the positive-norm branch in the slow region (left panel of Fig.\ref{fig:superradiant-dispersion}) to the negative branch in the fast region (right panel) occurs by tunneling through a local mass gap, a smoother transition implies a spatially wider mass gap and, thus, smaller values of the transmission and amplification coefficients. However, as these microscopic details do not qualitatively alter the structure of the waves far away from the transition region nor the mechanism of the emission process, the results that we are going to present for the step-like model are good representatives of a more general class of smooth configurations, with the key advantage of maximizing the strength of the superradiant process.

Another technical advantage of working with a step-like interface is that a full scattering solution can be obtained by simply requiring the continuous matching of the two plane-wave decompositions \eqref{eq:inout-modes} on either side of the ergosurface $y=0$ and of their first derivatives along $y$. This provides a number of four conditions on the amplitudes, which exactly equals the one of the four amplitudes involved in each scattering solution, so the linear system is well determined with a unique solution.
Once an ingoing mode is chosen at a given $k_x$, solving the scattering problem involves in fact four  amplitudes, divided between outgoing and evanescent modes. These latter do not enter in the scattering matrix since they are not relevant in the asymptotic regions, still they are important near the interface and are involved in the matching conditions. 

\begin{figure}[t]
	\centering
	\includegraphics[width=\columnwidth]{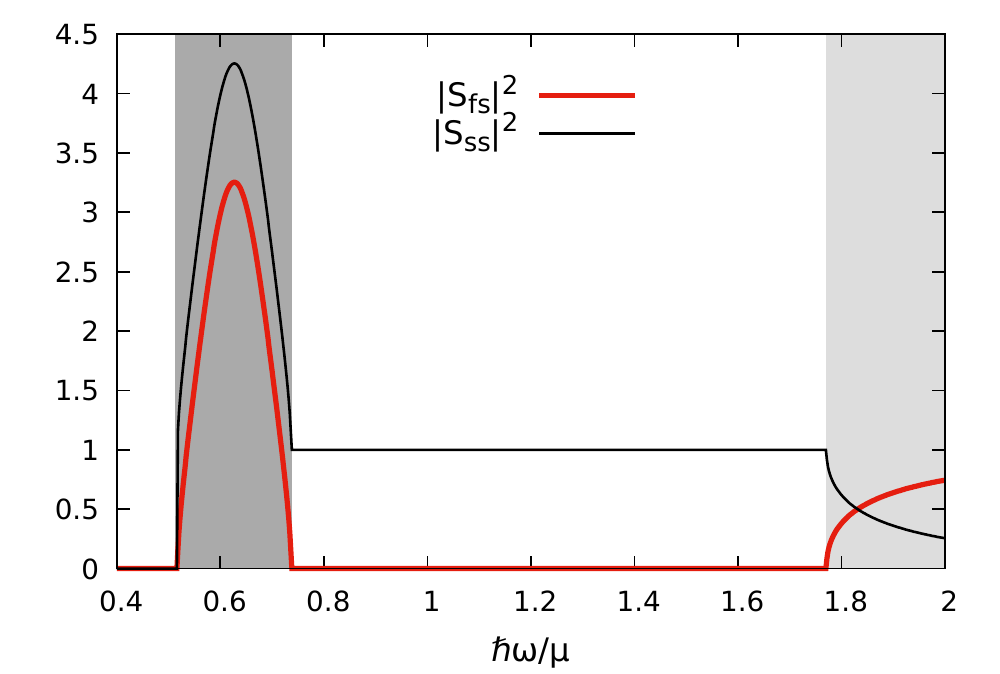}
	\caption{Reflection $|S_{ss}|^2$ (black thin line) and transmission $|S_{fs}|^2$ (red thick line) coefficients obtained with the matching of the modes at a step-like interface with $v_x^f=2.5\ c_s$ for an ingoing wave from the lower region with for $k_x=0.5\xi$. The dark and light grey regions correspond to the ones of Figure \ref{fig:superradiant-dispersion}: In particular, in the dark grey superradiant range, once can see that the amplified reflection coefficient goes well above one.}
	\label{fig:superradiant-coeffs-25}
\end{figure}

The first step to solve the problem consists of numerically finding the roots of the dispersion relations \eqref{eq:dispersion} in the two regions for the chosen $k_x$ and $\w$ and computing their group velocities  along $y$. Once we have selected the desired ingoing channel, for each region we keep the two physically relevant (outgoing or evanescent) roots and we numerically solve the corresponding linear problem for the mode amplitudes $\A_{I|out}$ and $\A_{I|ev}$ for $I=s,f$. The resulting values for the amplitudes of the propagating outgoing modes give the coefficients of the scattering matrix. The result of this procedure for an ingoing wave from the slower region $s|in$ in a regime where the condition \eqref{eq:superradiant-condition} for superradiant scattering is satisfied is illustrated in Figure \ref{fig:superradiant-coeffs-25}, where we plot the reflection $|S_{ss}|^2$ and the transmission $|S_{fs}|^2$ coefficients. 

These results can be understood in comparison with the dispersion relation plotted in Figure \ref{fig:superradiant-dispersion}. Energies below $\h\w/\mu\simeq0.5$ correspond to the mass gap and no ingoing mode is available for the scattering process. For $0.7\lesssim\h\w/\mu\lesssim1.75$, no travelling mode is available for transmission in the upper region, so reflection is total  ($S(k_x,\w)=[S_{ss}]$ with $|S_{ss}|^2=1$). For higher energies in the light gray region, positive-norm travelling modes are available in the upper region, so one has an ordinary scattering process with both reflection and transmission coefficients below one; for these frequencies the scattering matrix is unitary, i.e. $\eta=I$. 

The most interesting range lies between $0.5 \lesssim \h\w/\mu \lesssim 0.7$ and is indicated with the dark grey shading. Here, negative-norm outgoing modes are available in the upper region: The scattering matrix is hence pseudo-unitary with $\eta=\sigma_3$ and, as expected, one has superradiant scattering with a reflection coefficient $|S_{ss}|^2$ going above one. The extra energy is provided by the negative energy that is transmitted into the negative-norm mode in the upper region: As expected from pseudo-unitarity, the difference $|S_{ss}|^2-|S_{fs}|^2$ is constant and equal to $1$.

Note that the solution of the scattering problem with the ingoing (negative-norm) wave from the fast region gives the same value of the reflection coefficient in the superradiant interval, meaning that the magnitude of superradiant amplification does not depend on the direction in which the ergosurface is crossed. This is a consequence of the pseudo-unitarity of the scattering matrix that implies
\begin{equation}\label{eq:transmission-symmetry}
	|S_{fs}|^2=|S_{sf}|^2.
\end{equation}

\begin{figure}[t]
	\centering
	\includegraphics[width=\columnwidth]{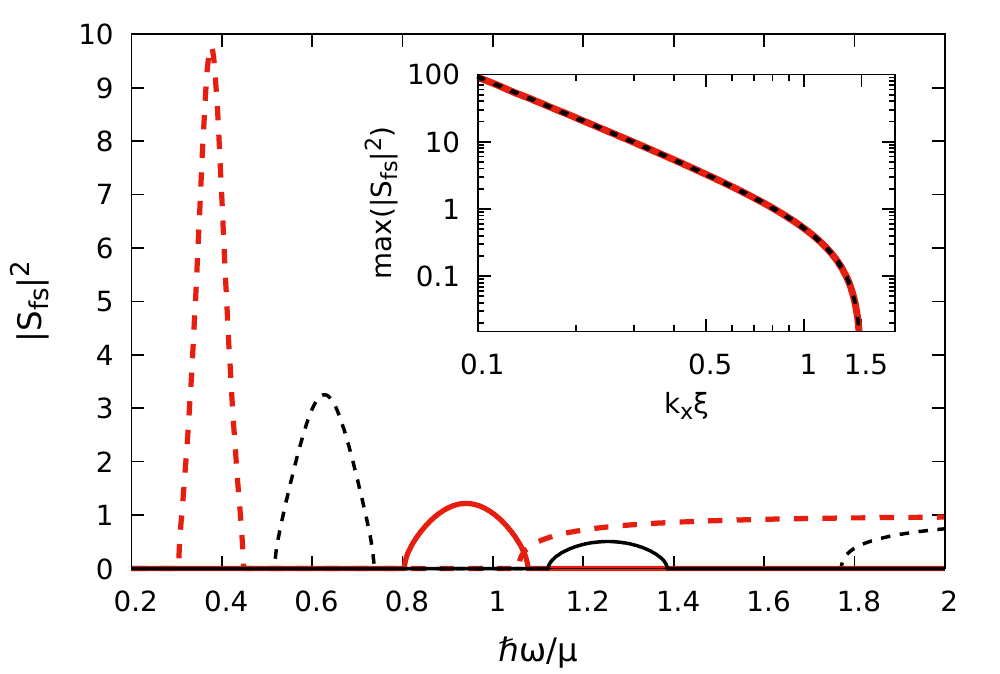}
	\caption{Main plot: Incident-energy-dependent transmission coefficient for an ingoing wave from the lower region and a fixed $v_x^f=2.5\ c_s$. The different (thick red dashed, thin black dashed, thick red solid, think black solid) curves correspond to different values of the transverse incident wavevector $k_x\xi=0.3,\ 0.5\ ,0.75,\ 1$. Inset: Log-log plot of the $k_x$-dependence of the transmission maximum. The solid red line shows the numerical data, the superimposed black dashed line is a fit of the $\frac{\alpha}{k_x^2}-\beta$ form, which reproduces the data very accurately.}
	\label{fig:superradiant-multple-k}
\end{figure}

The momentum-dependence of superradiant scattering can be investigated by solving the scattering problem at fixed $A_x^f$ for different values of the transverse momentum $k_x$. In Figure \ref{fig:superradiant-multple-k} one can see that the maximum of the transmission coefficient decreases for increasing $k_x$, while the width of the superradiant region widens. For even higher values of $k_x$, the maxima continue to decrease but the superradiant region shrinks again and eventually vanishes when the threshold \eqref{eq:dispersive-threshold} of dispersive suppression is reached. The $k_x$-dependence of the transmission maximum is plotted in the inset of Figure \ref{fig:superradiant-multple-k}. The maximum transmission, and hence the maximum superradiant amplification, is accurately reproduced by a $\alpha/k_x^2-\beta$ law with constant $\alpha,\beta>0$, the latter coefficient ensuring that the transmission coefficient vanishes at the dispersive suppression threshold, here approximately at $k_x\xi=1.5$.

\begin{figure}[t]
	\centering
	\includegraphics[width=\columnwidth]{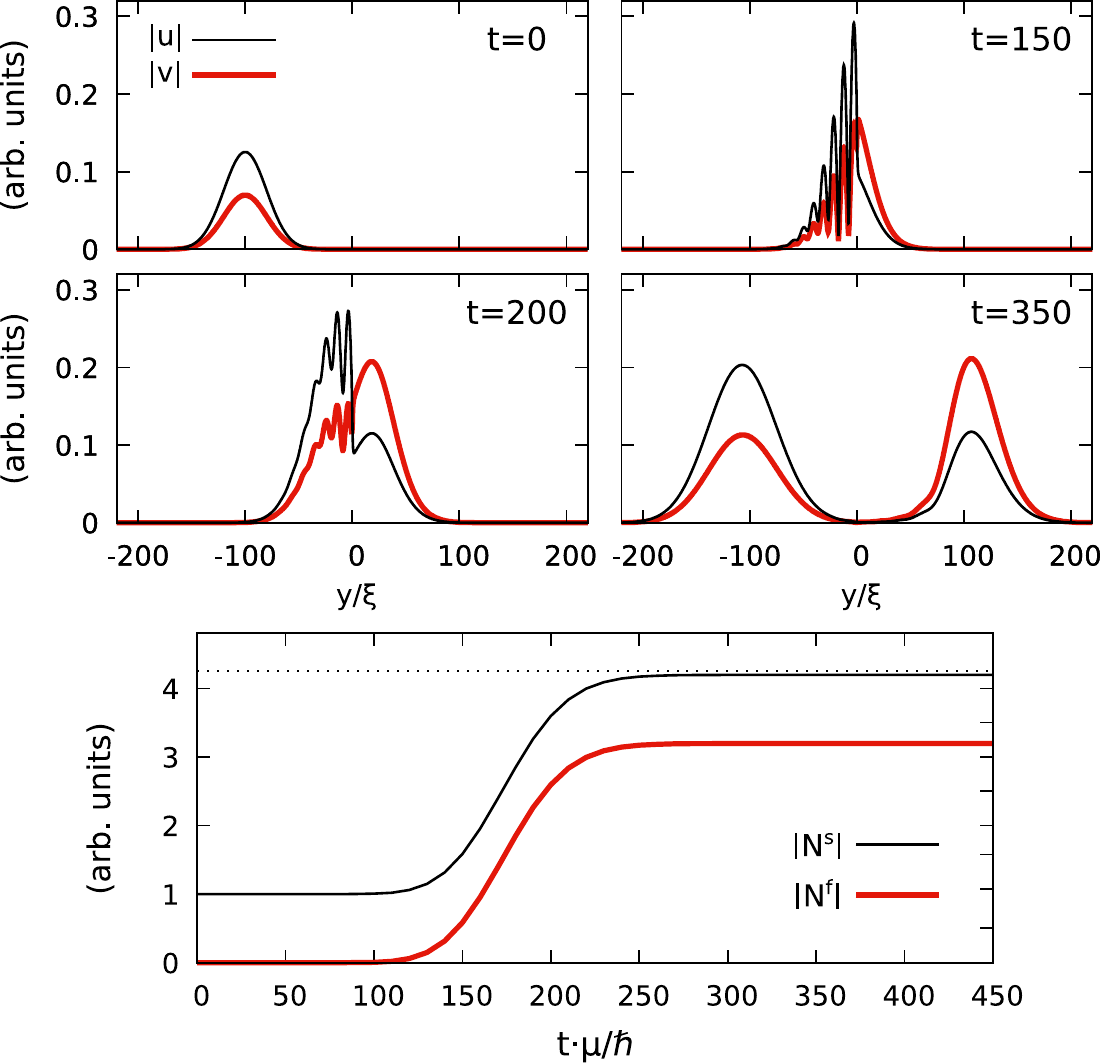}
	\caption{Top and central panels: Snapshots at different times $t\mu/\h=0, 150, 200, 350$ of the time evolution of a superradiant scattering process on an step-like ergosurface separating regions with $v_x^s=0$ and $v_x^f=2.5\ c_s$. The incident wavepacket is Gaussian with a transverse momentum $k_x\xi=0.5$ and a longitudinal momentum centered around $k_y^{in}\xi=0.32$, that is $\h\w=0.63\ Mc_s^2$. It is initially located in space around $y_0=-100\ \xi$ with $\sigma=20\ \xi$. The black thin lines are the modulus of the first component $|U|$ of the Bogoliubov spinor, the red thick ones the modulus of the second one $|V|$. The amplification can measured by computing the total Bogoliubov norm \eqref{eq:packet-bogo-norm} in the two regions, whose time dependence is shown in the lower panel, with the black thin line being the modulus of the norm $|N^s|$ in the slow region and the red thick line the one $|N^f|$ in the fast region. The amplification coefficient is compatible with the maximum value found in Figure \ref{fig:superradiant-coeffs-25} and indicated here by the horizontal dotted line.}
	\label{fig:superradiant-timedep}
\end{figure}

An intuitive illustration of the physics underlying this scattering matrix can be obtained by performing a numerical evolution in time of the Bogoliubov problem of the whole configuration starting from a wavepacket in the slow region with group velocity directed towards the ergosurface. We performed such simulation with a third-order Runge--Kutta algorithm for the same parameters used for Figure \ref{fig:superradiant-coeffs-25} and starting from a Gaussian wavepacket localized within the dark grey superradiant region and peaked around the frequency of the maximum of the reflection coefficient. Snapshots of the evolution of the two components $u,v$ of the Bogoliubov spinor at different times are shown in the upper and central plots of Figure \ref{fig:superradiant-timedep}. 

The evolution of the norm of the packets is summarized in the lower plot where we show the time-dependence of the Bogoliubov norm in the two regions, obtained with the integral
\begin{equation}\label{eq:packet-bogo-norm}
	N^{s,f}(t)=\int_{s,f} \mathrm{d}y\;\left(|U(t,y)|^2-|V(t,y)|^2\right),
\end{equation}
where $U(t,y)$ and $V(t,y)$ are the time-dependent components of the Bogoliubov spinor, the subscript $s$ ($f$) on the integral indicating that the integration is performed on the $(-\infty,0)$ interval (the $(0,+\infty)$ interval). One can see that the initial $s|in$ packet has a positive norm, since the $u$ component of the spinor is larger than the $v$ one, while the transmitted $f|out$ packet has a negative norm. As a result the reflected $s|out$ packet has a norm approximately $4$ times larger than the ingoing one, in agreement with the maximum amplification factor we had found in Figure \ref{fig:superradiant-coeffs-25}. We have checked (not shown) that different choices of the incident wavevector $k_x$ and energy $\h\w$ lead to the other behaviours discussed above.

\section{Quantum description and spontaneous pair production}
\label{sec:quantization}

In all our discussion so far, we treated the fluctuations in the Bogoliubov field as a classical field, relying on the fact that superradiant scattering is a classical field effect. From both quantum optics~\cite{QuantumOptics} and gravitational physics, it is in fact known that amplification generally comes in pair with spontaneous emission. A quantized theory of the field undergoing amplified superradiant scattering in a rotating black hole spacetime predicts the appearance of a spontaneous quantum pair production in the modes responsible for superradiance~\cite{ford1975quantization}. These processes can be physically understood as a superradiant amplified scattering of vacuum fluctuations, that end up populating the opposite-normed outgoing modes as real quanta of radiation. 
Taking inspiration from the input-output approach developed in~\cite{recati2009bogoliubov} to describe the related phenomenology of Hawking radiation, a general theory of spontaneous superradiant emission can be obtained by extending scattering matrix approach of the previous Section to the quantum context. This is the subject of the present Section.

Fluctuations around a stationary state of a BEC can be described by a bosonic quantum field $\widehat{\Psi}=\Psi_0+\widehat{\delta\Psi}$, where the order parameter $\Psi_0$ describing the condensate continues to be treated as a classical field and the quantum behaviour is encoded in the fluctuations, namely in the non-condensed fraction. This quantum field can be derived in a rigorous way~\cite{castin2001bose} by applying a second quantization
procedure to the classical field: After choosing a suitable basis for the field modes to isolate the condensate one, annihilation operators are associated to positive-norm modes and creation operators to negative-norm ones. In our case, convenient bases are provided by the scattering modes: For each $k_x$ and $\w$, one can construct a basis of the field by taking the scattering solutions corresponding to a single non-vanishing ingoing wave, this is called the basis of ingoing modes. Another equally good basis is instead provided by the scattering solutions displaying only one non-vanishing outgoing mode, this will be the basis of the outgoing modes. Both these bases can be used to quantize the field.

For the sake of brevity, we restrict our discussion here on the most interesting frequency components located within in the superradiant range and comment on the other regimes when needed. In this regime, the quantum field $\widehat{\delta\Psi}^{SR}(y)$ can be expressed in the basis of ingoing scattering modes and their corresponding operators $\hat a_{s,f}(k_x,\w)$ as
\begin{equation}\label{eq:field-quantization}
\begin{split}
	\widehat{\delta\Psi}^{SR}(y)=\int_0^{k_x^{max}}\mathrm{d}k_x \int_{\w_{min}}^{\w_{max}}\mathrm{d}\w\ [U_{s|in}(y)\hat a_{s} + V^*_{s|in}(y)\hat a_{s}^\dag\\
	+ U_{f|in}(y)\hat a_{f}^\dag + V^*_{f|in}(y)\hat a_{f}]\,.
\end{split}
\end{equation}
Here, $U_{I|in}(y)$ and $V_{I|in}(y)$ are the Bogoliubov components of the ingoing scattering modes, $\w_{min}=\B(k_x)/\h$ and $\w_{max}=v^f_xk_x-\B(k_x)/\h$ indicate the limits of the superradiant frequency interval, and the dependence on $k_x$ and $\w$ of the spinor components and of the operators is kept implicit to improve readability. Note in particular the exchange of creation and annihilation operators in the second line, due to the negative norm of the transmitted mode in the fast region. An analogous expression of the quantum field can be written in terms of outgoing scattering modes, whose annihilation operators we instead indicate with $\hat b_{s,f}(k_x,\w)$.

A relation between the two sets of operators is given by the input-output relation \eqref{eq:inputoutput}, that, for frequencies in the superradiant interval, takes here the following form
\begin{equation}\label{eq:quantum-io-sr}
	\begin{pmatrix}
		\hat{b}_s(k_x,\w)\\ \hat{b}_f^\dag(k_x,\w)
	\end{pmatrix}
	=S(k_x,\omega)
	\begin{pmatrix}
		\hat{a}_s(k_x,\w)\\ \hat{a}_f^\dag(k_x,\w)
	\end{pmatrix}
\end{equation}
where the modes in the fast region appear via their creation operators as a consequence of their negative norm. A similar scattering matrix mixing creation and destruction operators underlies the theory of Hawking radiation developed in~\cite{recati2009bogoliubov}. In contrast to the three modes involved in the Hawking emission~\cite{busch2014}, the structure of the present superradiant process is simpler and only involves a pair of opposite-norm modes. This will be beneficial in view of observing clean entanglement features in the superradiant emission~\cite{finazzi2014entangled}.

For incident frequencies in the ordinary scattering $\h\w>-\h v^f_xk_x +\B(k_x)$ range, the available modes in the fast region also have a positive norm, so the expression of the field does not show the exchange of creation and annihilation operators of the second line of \eqref{eq:field-quantization} and the input-output relation is given by the (unitary) scattering matrix as 
\begin{equation}\label{eq:quantum-io-os}
	\begin{pmatrix}
		\hat{b}_s(k_x,\w)\\ \hat{b}_f(k_x,\w)
	\end{pmatrix}
	=S(k_x,\omega)
	\begin{pmatrix}
		\hat{a}_s(k_x,\w)\\ \hat{a}_f(k_x,\w)
	\end{pmatrix}.
\end{equation}
The situation is further simplified in the case of total reflection where the input-output relation reduces to a scalar equation 
\begin{equation}
    \hat b_I(k_x,\w)=S(k_x,\w)\hat a_I(k_x,\w)
    \label{eq:quantum_io_scal}
\end{equation}
with $|S|=1$.
In these two last cases, the representations in terms of input and output modes of the quantum field are unitarily equivalent. In the superradiant case of equation \eqref{eq:quantum-io-sr}, instead, the mixing of creation and annihilation operators indicates the inequivalence of the two representations: As discussed for example in~\cite{jacobson2005introduction}, this is the basic mathematical origin of particle creation effects in quantum field theories in curved spacetimes, including Hawking radiation and our spontaneous superradiant emission.

\begin{figure}[t]
	\centering
	\includegraphics[width=\columnwidth]{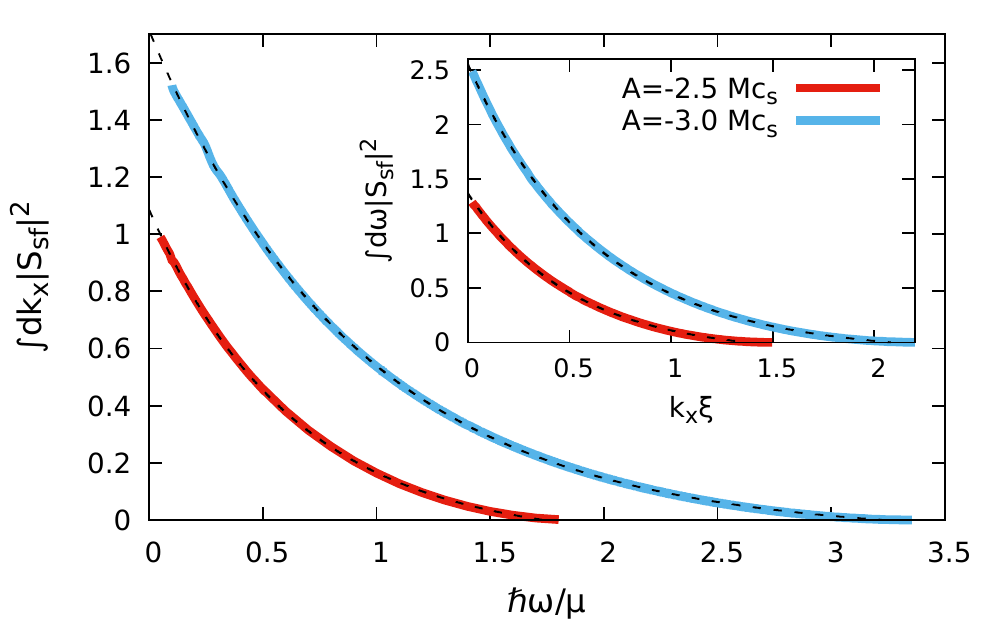}
	\caption{Main plot: Frequency-dependent spectrum of the spontaneous quantum superradiant emission after integration over transverse momentum $k_x$. 
	Inset: Momentum-dependent spectrum after integration over frequencies $\w$. The different curves correspond to different values of the synthetic vector potential in the fast region, such that $v_x^f=2.5\;c_s$ (solid red) and $v_x^f=3\;c_s$ (solid blue). In all curves, the synthetic vector potential (and hence the velocity) vanishes $A_x^s=0$ in the slow region.
	The black dashed lines are fits of the numerical curve with a shifted exponential law of the form $f(\w)=\a\exp(-\b \w)-\gamma$ (and analogous expression with $k_x$ instead of $\w$) with $\a,\b,\gamma$ positive parameters. Such fits are found to describe very accurately the numerical data. }
	\label{fig:spontaneous-emission-spectrum}
\end{figure}

Let us focus on the flux of outgoing phonons spontaneously emitted into the slow region. Using the input-output relation, one obtains
\begin{equation}\label{eq:spontaneous-emission}
	\braket{\hat b_{s}^\dag \hat b_{s}}=|S_{ss}|^2\braket{\hat a_s^\dag \hat a_s}+|S_{sf}|^2\left(1+\braket{\hat a_f^\dag \hat a_f}\right),
\end{equation}
where the dependence of all the quantities on $k_x$ and $\w$ is understood and the $1$ in the last term comes from the commutation of the negative-norm operators. Because of this constant term, the flux of the outgoing modes remains finite even if the initial populations of the ingoing modes are zero (as it happens, e.g., at zero temperature), and is given by the $|S_{sf}(k_x,\w)|^2$ matrix element between opposite norm operators. 

Because of the symmetry \eqref{eq:transmission-symmetry} of the scattering matrix, the same conclusion holds for the outgoing mode in the fast region, which also displays a flux 
\begin{equation}
\braket{\hat b_{f}^\dag \hat b_{f}}=|S_{ff}|^2\braket{\hat a_f^\dag \hat a_f}+|S_{fs}|^2\left(1+\braket{\hat a_s^\dag \hat a_s}\right)\end{equation}
which is non-zero even for vanishing initial populations.
On the other hand, for an ordinary scattering the input-output scattering matrix \eqref{eq:quantum-io-os} and \eqref{eq:quantum_io_scal} do not involve any creation operator, so there will be no spontaneous emission. 

For each $k_x$, the $\w$-dependent spectrum of the spontaneous emission is hence given by the superradiant \textit{bump} of the transmission coefficient visible in the dark grey region of Figure \ref{fig:superradiant-coeffs-25}. The overall emission spectrum can be obtained by restricting to superradiant frequency ranges and integrating the transmission coefficient $|S_{sf}(k_x,\w)|^2$ in $k_x$. The result of such a calculation using the prediction for the scattering matrix elements obtained in Section \ref{sec:scattering-matching} is shown in the main plot of Figure \ref{fig:spontaneous-emission-spectrum} for two values of the vector potential $A_x^f$ in the fast region, that is the condensate speed. 
In both cases, the numerically obtained spectrum shows a smooth decrease of the emission with energy $\w$. 

Remarkably, for each value of $A_x^f$, this dependence can be fitted with great precision with a shifted decreasing exponential of the form $f(\omega)=\alpha \exp(-\beta \omega) -\gamma$. The main effect of the constant shift $\gamma$ is to enforce the emission to vanish above the dispersive threshold \eqref{eq:dispersive-threshold}. 
The smooth and regular shape of this spectrum is qualitatively different from the one of the Hawking emission, whose spectrum was found to diverge according to a $\w^{-1}$ thermal law at small frequencies for all values of the surface gravity~\cite{Macher1,Macher2}. Whereas this low-$\w$ divergence of the Hawking spectrum is preserved even for the formally infinite surface gravity of a step-like acoustic horizon~\cite{recati2009bogoliubov}, no thermal behavior is ever expected for the superradiant emission and the spectrum maintains a regular shape even for smoother transitions. 

For the sake of completeness,  the transverse momentum distribution of the spontaneous emission after integrating $|S_{sf}(k_x,\w)|^2$ in $\w$ is shown in the inset of Figure \ref{fig:spontaneous-emission-spectrum}. Also in this case, the spectrum shows an analogous shifted decreasing exponential behaviour, showing that the emission is larger in the smaller transverse momenta.

\section{Signatures of the emission on the correlation functions}
\label{sec:corr}

One of the key features of analog Hawking radiation are the quantum correlations that exist between the Hawking particle propagating away from the horizon in the outward direction and its partner propagating in the inward direction inside the black hole. Such correlations directly translate into a specific feature in the correlation function of density fluctuations on either side of the horizon~\cite{balbinot2008nonlocal,carusotto2008numerical} and provided the smoking gun of Hawking emission in the recent experiments~\cite{steinhauer2016observation,de2019observation}.

Given the close physical analogy of our scattering matrix description \eqref{eq:quantum-io-sr} of superradiant emission to the one of analog Hawking emission in~\cite{recati2009bogoliubov}, it is natural to expect that similar correlations should appear in the superradiant emission.
This is formally captured by the non-vanishing anomalous correlation
\begin{equation}
	\braket{\hat b_{s}^\dag(k_x,\w) \hat b_{f}^\dag(k_x,\w)}=S_{sf}^*S_{ff},
\end{equation}
between the opposite-norm modes propagating away from the ergosurface in the two regions. 

In the experimental observation of analog Hawking emission in a BEC \cite{steinhauer2016observation,de2019observation}, these correlations were extracted from a measurement of density-density correlations in position space following the proposal in~\cite{balbinot2008nonlocal,carusotto2008numerical}. In the next Subsection \ref{sec:space-corr} we will explore this same route in the superradiant context. 
In the following Subsection \ref{sec:momentum-corr} we will take inspiration from~\cite{boiron2015quantum,fabbri2018momentum} to investigate a alternative scheme based on momentum-space correlations which is expected to provide a neater signature of quantum superradiant emission.

\subsection{Density-density correlations in position space}
\label{sec:space-corr}
As usual, the correlation function of position-space density fluctuations is expressed as the normal ordered product
\begin{equation}
	\begin{split}
	G^{(2)}(\mathbf{r},\mathbf{r}')=&\braket{\widehat\Psi^\dag(\mathbf{r})\widehat\Psi^\dag(\mathbf{r}')\widehat\Psi(\mathbf{r}')\widehat\Psi(\mathbf{r})}\\
	&- \braket{\widehat\Psi^\dag(\mathbf{r})\widehat\Psi(\mathbf{r})} \braket{\widehat\Psi^\dag(\mathbf{r}')\widehat\Psi(\mathbf{r}')}.
	\end{split}
	\label{eq:G2_spatial}
\end{equation}
Expanding the quantum field in terms of the condensate fraction plus small fluctuations, $\widehat\Psi=\Psi_0+\widehat{\delta\Psi}$, and
keeping only the terms of the second order in the fluctuation field one obtains
\begin{equation}\label{eq:correlation-workedout}
	\begin{split}
	G^{(2)}(\mathbf{r},\mathbf{r}')&= \braket{\widehat{\delta\Psi}(\mathbf{r}')\widehat{\delta\Psi}(\mathbf{r})}+\braket{\widehat{\delta\Psi}^\dag(\mathbf{r}')\widehat{\delta\Psi}(\mathbf{r})} \\
	&+\braket{\widehat{\delta\Psi}^\dag(\mathbf{r})\widehat{\delta\Psi}(\mathbf{r}')} +\braket{\widehat{\delta\Psi}^\dag(\mathbf{r})\widehat{\delta\Psi}^\dag(\mathbf{r}')}.
	\end{split}
\end{equation}
Given the translational symmetry of our problem, the correlation function only depends on $(x-x',y,y')$. We can then expand the one-time correlation function in its $\w$ and $k_x$ components as
\begin{equation}\label{eq:correlations-components}
	G^{(2)}(\mathbf{r},\mathbf{r}') = \int_{-\infty}^\infty \frac{\mathrm{d}k_x}{2\pi}\,e^{ik_x(x'-x)}\, \int_{-\infty}^\infty \mathrm{d}\w\ G^{(2)}(k_x,\w,y,y')
\end{equation}
and make use of the scattering matrix formalism developed in the previous Section to evaluate each $k_x,\omega$ component $G^{(2)}(k_x,\omega,y,y')$. 

To this purpose, we can make use of the expansion \eqref{eq:field-quantization} of the field in terms of outgoing modes and then use the input-output relation \eqref{eq:quantum-io-sr}.
Focusing on the case in which $y<0$ and $y'>0$ and considering points sufficiently far from the interface so that one can neglect the evanescent modes, one can write the field in each region as
\begin{equation}
	\begin{split}
	\widehat{\delta\Psi}_{SR}(k_x,&\w,y<0)=\\
	&\tld U_{k_{s|out}}e^{ik_{s|out}y}\;\hat b_s + \tld V_{k_{s|out}}^*e^{-ik_{s|out}y}\;\hat b_s^\dag\\
	&+ \tld U_{k_{s|in}}e^{ik_{s|in}y}\;\hat a_s + \tld V_{k_{s|in}}^*e^{-ik_{s|in}y}\;\hat a_s^\dag
	\end{split}
\end{equation}
and
\begin{equation}
	\begin{split}
	\widehat{\delta\Psi}_{SR}(k_x,&\w,y'>0)=\\
	&\tld U_{k_{f|out}}e^{ik_{f|out}y'}\hat b_f^\dag  + \tld V_{k_{f|out}}^*e^{-ik_{f|out}y'}\hat b_f\\
	&+ \tld U_{k_{f|in}}e^{ik_{f|in}y'}\hat a_f^\dag + \tld V_{k_{f|in}}^*e^{-ik_{f|in}y'}\hat a_f.
	\end{split}
\end{equation}
Here, for compactness, the Bogoliubov components with the tildes also include the normalization factors appearing in equation \eqref{eq:inout-modes}, and we omitted the $k_x$ and $\w$ dependence in the right-hand side.

Starting from these expressions one can compute the two-point correlators on the vacuum of the ingoing modes by substituting the $\hat b$ operators via the input-output relation \eqref{eq:quantum-io-sr} and keeping only the quantum fluctuations terms coming from commutators. Only terms proportional to the product of two outgoing Bogoliubov amplitudes or to the product of one ingoing and one outgoing amplitude remain. The exponential factors of the first kind of terms have a phase that can be stationary for $y<0$ and $y'>0$, while the the ones of the second kind of term will be fast oscillating in this quadrant. We can hence neglect these terms to compute the correlations at sufficiently long distance from the interface. 

With these considerations and using the pseudo-unitarity of the scattering matrix, we obtain the expression

\begin{equation}\label{eq:inout-correlations}
	\begin{split}
		G_{SR}^{(2)}(k_x,\w,y<0,y'>0)= \frac{R_{k_{s|out}}(k_x,\w) R_{k_{f|out}}^*(k_x,\w)}{2\pi\sqrt{|v_{s|out} v_{f|out}|}}\\ 
		\times S_{sf}(k_x,\w)S_{ff}^*(k_x,\w)\ e^{i[k_{s|out}(k_x,\w)y-k_{f|out}(k_x,\w)y']} + c.c.\;,
	\end{split}
\end{equation}
for the density correlation function due to the quantum superradiant emission. Here, $R_{k_{I|out}}:=U_{k_{I|out}}+V_{k_{I|out}}$ (with $I=s,f$) are shorthands for suitable combinations of the Bogoliubov components of the plane wave expansion \eqref{eq:inout-modes}.  Since this equation does not include the evanescent mode and we neglected the fast oscillating terms, the result will be quantitatively reliable only sufficiently far from the interface.
Expression \eqref{eq:inout-correlations} is formally analogous to the one obtained for Hawking radiation in \cite{recati2009bogoliubov}, with the key addition of the transverse degrees of freedom encoded in the $k_x$ dependence. This seemingly innocuous addition actually introduces serious complications in the real-space correlation pattern, as we are going to see in the following. This was noted in~\cite{Dudley18}, while studying the effect of a finite transverse momentum of the excitations on the density correlations due to Hawking emission.

Since we saw that quantum pair production in our setup only occurs for frequencies in the superradiant window, this frequency range is the only source of correlations between the slow and fast regions. An analogous calculations for the ordinary scattering regime of \eqref{eq:quantum-io-os} shows in fact that in this case all contributions to correlation between opposite sides of the ergosurface vanish because of the unitarity of the scattering matrix. 

\begin{figure}[t]
	\centering
	\includegraphics[width=\columnwidth]{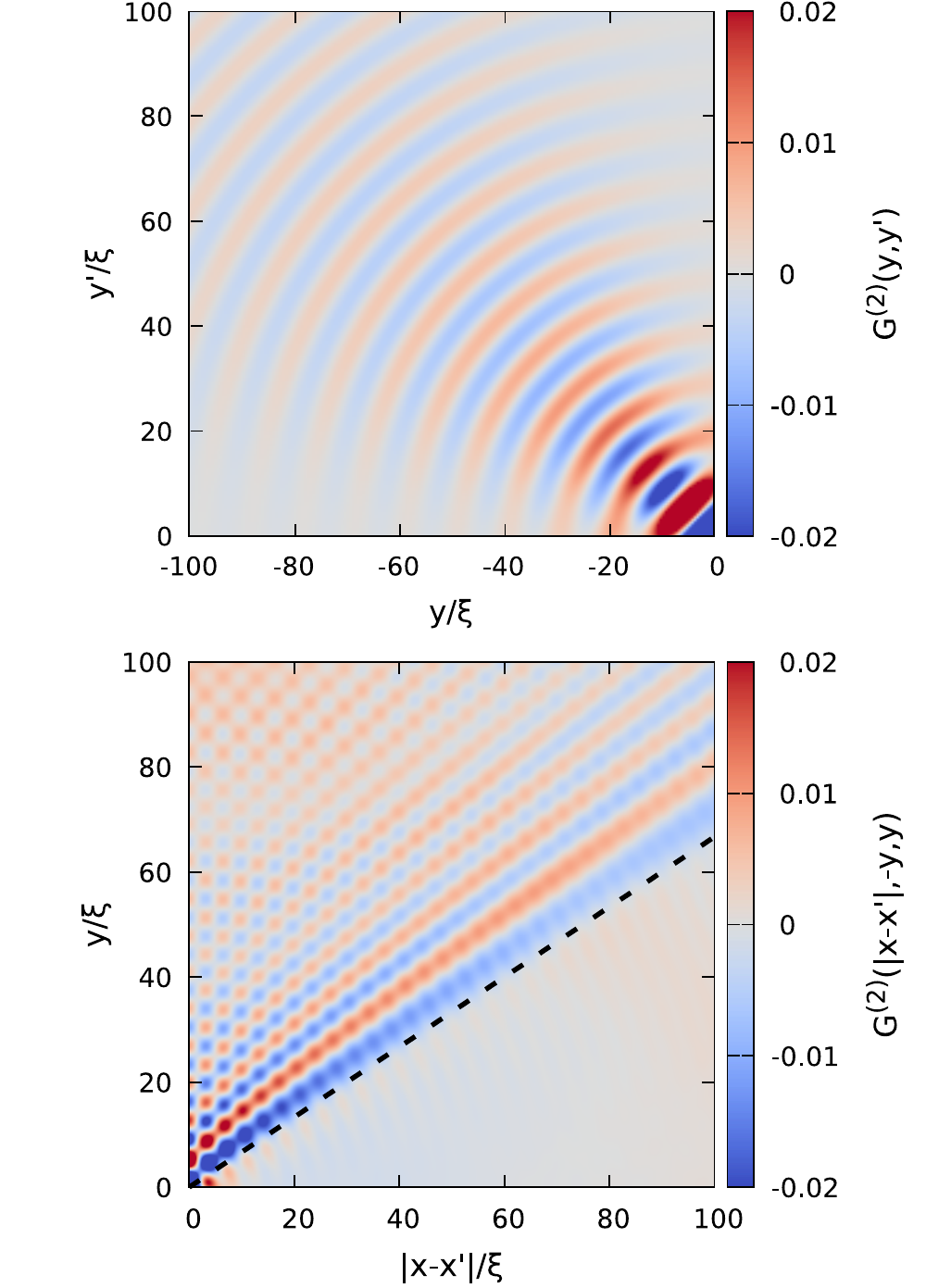}
	\caption{Colorplots of the position space correlation function $G^{(2)}$ of density fluctuations for a pair of points $(x,y)$ and $(x',y')$ located respectively in the slower ($y<0$) and faster ($y'>0$) regions. Because of the translational invariance along $x$, the correlation function only depends on $|x-x'|$. The upper plot shows the $y,y'$ dependence of $G^{(2)}$ for $x-x'=0$, while the lower one shows its dependence on $x-x'$ and $y$ for symmetrically located $y'=-y$. The black dashed line in the lower plot has a slope $\left[(v_x^f/c_s)^2-4\right]^{-1/2}$ and well reproduces the boundary of the main contributions to the correlation function. The value of the synthetic vector potential in the faster region is taken so that $v^f_x=2.5\ c_s$, while in the slower region $v_x^s=0$. The numerical calculation was performed using the scattering matrix approach for a system initially in the ground state.}
	\label{fig:density-correlations}
\end{figure}

The full spatial dependence of density correlations on opposite sides of the ergosurface can be evaluated by inserting in \eqref{eq:inout-correlations} the scattering matrix elements obtained in the previous Section and then performing the integrals over $k_x$ and $\omega$. Notice that, because of the dispersive threshold \eqref{eq:dispersive-threshold}, the $k_x$ integration domain in \eqref{eq:correlations-components} is actually restricted to the finite interval $[-k_x^{max},k_x^{max}]$.
An example of the result of such a calculation is shown in Figure \ref{fig:density-correlations}. Since the position space correlation function $G^{(2)}$ depends on the three variables $x-x',y,y'$, a full three-dimensional plot is impractical and suitable cuts need being chosen for the plots. 

The upper panel shows $G^{(2)}$ as a function of the $y,y'$ coordinates for $x-x'=0$. The plot displays an oscillating behaviour that extends throughout the $y,y'$ plane, with an amplitude that is maximal along the diagonal $y=y'=0$ and decreases while approaching the $y=0$ and $y'=0$ axis. The complexity of this plot is to be contrasted with the simple structure of the correlation features found for one-dimensional black holes configurations. In \cite{recati2009bogoliubov} the different features of this case were explained in terms of the group velocities of the involved modes; in particular, strongest correlation were found at points for which $y'=(v_{f|out}/v_{s|out})y$, where the two group velocities are the ones of the partner modes in the emission. For the (approximately) linear dispersion relation involved in the Hawking processes, the (approximately) constant group velocity results in the correlations being peaked along straight lines. In contrast, in the present two-dimensional superradiant case, one can see from the dispersion relations of Figure \ref{fig:superradiant-dispersion} that for any given $k_x$, increasing the frequency $\w$ across the superradiant region makes the $y$-component of the group velocity $v_{s|out}$ in the slow region to grow from zero to some maximum value, while the one $v_{f|out}$ of the fast region decreases from the same maximum value to zero. Hence the slope $v_{f|out}/v_{s|out}$ of the lines on which one expects correlation extrema at the different frequencies varies from $-\infty$ to $0$. This means that the correlation signal will span the whole $(y<0,y'>0)$ quadrant, as we indeed observe in the Figure.
The oscillating pattern can instead be ascribed to the difference between the outgoing momenta on the two sides~\cite{grivsins2016theoretical}. In agreement with this interpretation, the characteristic wavelength of this pattern is found (not shown) to decrease for an increasing strength of the synthetic vector potential $|A^f_x|$, and thus of the relative flow speed. 

The complexity of the position space correlation function is confirmed in the lower panel of Figure \ref{fig:density-correlations}, where we show another cut of the density correlations to highlight the dependence on the $x$ direction; in particular correlations at equal distances from the ergosurface $y=-y'$ are shown. Also in this case, correlations extend for a large area of the $y=-y',x-x'$ plane, the only remarkable feature being the boundary on the largest distance $|x-x'|$ that the main contributions to the correlation function
can reach for each value of $y$. Also the slope of this boundary can be understood in terms of the group velocities of the partner outgoing modes on the two sides of the interface. Since we are considering symmetrically located points $y=-y'$, the main contributions to $G^{(2)}$ come from couples of modes that have the same modulus $|v_{g,y}|:=|v_{s|out}|=|v_{f|out}|$ of the group velocity along $y$; this happens, at fixed $k_x$, for $\w=v_x k_x/2$, as can be understood by looking at Figure \ref{fig:superradiant-dispersion}.

Analogously to the other cut of $G^{(2)}$ we just discussed, extrema of the correlations can be expected to lay on lines $y=m(k_x)|x-x'|$ of slopes $m(k_x):=|v_{g,y}|/|v_{g,x}^f-v_{g,x}^s|$, where in the denominator the group velocities along $x$ of the involved modes appear. This ratio can be analytically computed starting from the dispersion relations \eqref{eq:dispersion}, and the slope of the main contributions to $G^{(2)}$ at each $k_x$ turns out to be
\begin{equation}
    m(k_x)=\left|\frac{q(k_x)}{\h k_x}\frac{\sqrt{Mq(k_x)-\h^2k_x^2-2M^2c_s^2}}{M(v_x^f)^2-2q(k_x)}\right|,
\end{equation}
where $q(k_x):=\sqrt{\h^2(v_x^f)^2k_x^2+4M^2c_s^4}$. This quantity reduces, for small $k_x$, to the non-dispersive result $m(k_x\to 0)=\left[(v_x^f/c_s)^2-4\right]^{-1/2}$, corresponding to the black dashed line in the lower panel of Figure \ref{fig:density-correlations}. For growing $k_x$, the slope increases starting from this value and diverges for some $k_x<k_x^{max}$, consistently with the non-trivial correlations we observe everywhere above the dashed line. For higher transverse momenta $m(k_x)$ decreases instead from infinity to zero, while approaching $k_x^{max}$; the much smaller correlations below the dashed line stem hence from the fact that superradiant emission is dominated by the lower transverse momenta, as can be seen from the inset of Figure \ref{fig:spontaneous-emission-spectrum}.
Also in this case, the oscillating pattern is related to the difference between the outgoing momenta on the two sides. In addition to the stripes visible in upper panel, the lower panel shows a checkerboard-like pattern due to simultaneous presence of two emission channels with the same slope in the $y,|x-x'|$ plane.

Even though theoretically intriguing, these features of the correlation function do not appear to be experimentally as easy to detect as were the Hawking signatures in the one dimensional flows studied in~\cite{steinhauer2016observation,de2019observation}. For this reason, the next Subsection is devoted to the investigation of an alternative signature of spontaneous superradiant emission based on momentum-space correlations.

\subsection{Two-body correlations in momentum space}
\label{sec:momentum-corr}

As it was proposed in~\cite{boiron2015quantum,fabbri2018momentum} for analog Hawking radiation and experimentally adopted in~\cite{cayla2020hanbury} for a related Hanbury Brown and Twiss physics, two-body correlations in momentum space offer a promising different insight in the quantum fluctuations of many-body systems.
This physics is formally encoded in the correlation function \begin{equation}\label{eq:momentumcorr-general}
	G^{(2)}(k_x,k_y,k_y') =\langle \widehat{\delta\Psi}^\dag(k_y)\widehat{\delta\Psi}^\dag(k_y')\widehat{\delta\Psi}(k_y')\widehat{\delta\Psi}(k_y)\rangle,
\end{equation}
involving non-condensate atoms; here the dependence on $k_x$ of the field operators in the right hand side is understood. As it was discussed in~\cite{butera2020position}, the condensate lives in the zero-momentum state, so that its contribution can be filtered out with standard tools~\cite{cayla2020hanbury}.

In our geometry, a semi-analytical insight on momentum correlations can be obtained again using the scattering approach. For each value of the transverse momentum $k_x$ and the frequency $\w$, the dispersion relations in the two regions fix the values of the outgoing momenta $k_{s|out}(k_x,\w)$ and $k_{f|out}(k_x,\w)$ so that momentum correlations are only nonzero for these value of the momenta. Indicating with $k_y$ ($k_y'$) the momentum in the slow (fast) region, we can formally write
\begin{equation}\label{eq:momenta-corr-delta}
	\begin{split}
	G^{(2)}(k_x,\w,k_y,k_y')\propto &\delta(k_y-k_{s|out}(k_x,\w))\\
	&\times\delta(k_y'-k_{f|out}(k_x,\w)).
	\end{split}
\end{equation} 
Integrating over the frequency $\w$, for each transverse momentum $k_x$ the points of nonzero correlations will describe a line in the $(k_y,k_y')$ space. Joining all lines for different $k_x$, a surface in the $(k_y,k_y',k_x)$ space is found with specific geometric features. We expect that this surface will be a most recognizable signature of the quantum superradiant emission.

Along the lines of~\cite{butera2020position}, the quartic correlator \eqref{eq:momentumcorr-general} can be expressed through a Wick expansion in terms of products of second order correlators, 
\begin{equation}
	\begin{split}
	G^{(2)}&(k_y,k_y')=\abs{G^{(1)}(k_y,k_y')}^2\\
	&+ G^{(1)}(k_y,k_y) G^{(1)}(k_y',k_y') + \abs{A^{(1)}(k_y,k_y')}^2,
	\end{split}
\end{equation}
where the dependences on $k_x$ and $\w$ are implicit and the functions 
\begin{equation}
	G^{(1)}(k_y,k_y')=\langle \widehat{\delta\Psi}^\dag(k_y) \widehat{\delta\Psi}(k_y')\rangle
\end{equation}
and 
\begin{equation}
	A^{(1)}(k_y,k_y')=\langle \widehat{\delta\Psi}(k_y) \widehat{\delta\Psi}(k_y')\rangle
\end{equation}
are the usual normal and anomalous correlators.

In our planar superradiant configuration, expressions of these correlators at fixed $k_x$ and $\w$ for momentum values satisfying \eqref{eq:momenta-corr-delta} can be obtained following a procedure analogous to the one used for the position-space correlations. Assuming again the system to be initially in the ground state, we only keep the terms due to commutators, which, for each pair of $k_y,k_y'$ modes on the $s,f$ sides of the ergosurface, results in
\begin{subequations}
	\begin{align}
	\Big\lvert A^{(1)}(k_y&,k_y')\Big\rvert^2 = 2\Re\left(U_s^*V_s^*U_f V_f (S_{sf}^*S_{ff})^2\right)\nonumber
	\\&+ \left(\abs{U_s}^2\abs{V_f}^2 + \abs{V_s}^2\abs{U_f}^2\right)\abs{S_{sf}}^2\abs{S_{ff}}^2
	\\
	\Big\lvert G^{(1)}(k_y&,k_y')\Big\rvert^2 = 2\Re\left(U_s^*V_s^*U_fV_f (S_{sf}^*S_{ff})^2\right)\nonumber
	\\&+ \left(\abs{U_s}^2\abs{U_f}^2 + \abs{V_s}^2\abs{V_f}^2\right)\abs{S_{sf}}^2\abs{S_{ff}}^2
	\\
	G^{(1)}(k_y&,k_y)G^{(1)}(k_y',k_y') = \left(\abs{U_s}^2\abs{S_{sf}^2} + \abs{V_s}^2\abs{S_{ff}^2}\right)\nonumber
	\\ &\times\left(\abs{U_f}^2\abs{S_{ff}^2} + \abs{V_f}^2\abs{S_{sf}^2}\right).
	\end{align}
\end{subequations}
Here $U_I(k_y)=\delta(k_y-k_{I|out})U_{k_{I|out}}$, with $I=s,f$, and analogously for $V_I(k_y)$.

\begin{figure}[t]
	\centering
	\includegraphics[width=\columnwidth]{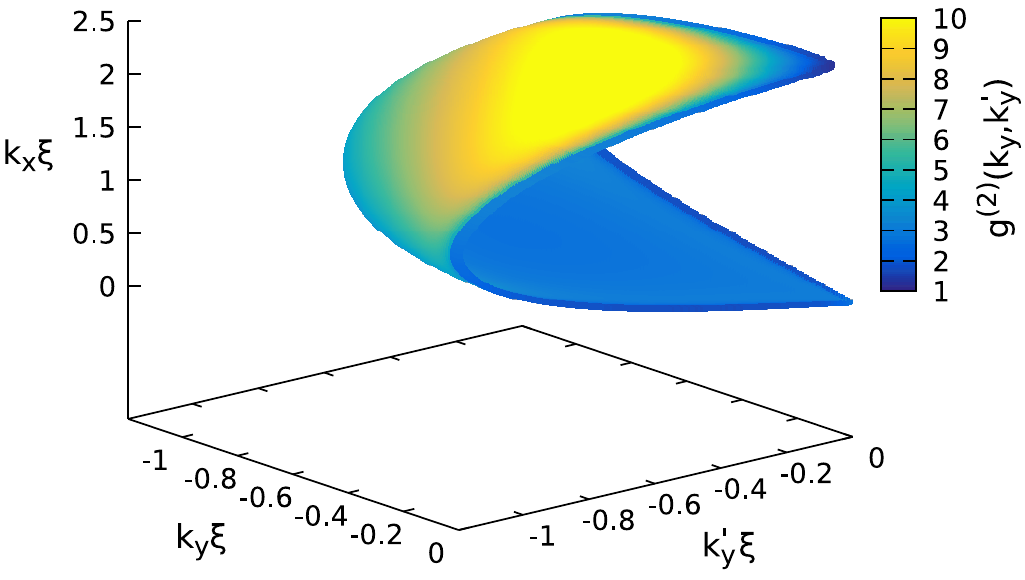}
	\caption{Plot of the normalized two body correlation function in momentum space for $v_x^f=3\;c_s$. For each value of $k_x$, non-trivial correlations are localized on a line indicating the momenta $k_y,k_y'$ of the modes involved in the superradiant emission. The union of all these lines describes the surface that is visible in the plot. The coloring of the surface indicates the strength of the normalized correlation function \eqref{eq:g2norm}. }
	\label{fig:correlation-surface}
\end{figure}

Both the locus of momentum space points satisfying  \eqref{eq:momenta-corr-delta} and the magnitude of two-body momentum correlations on these modes can be numerically computed within our scattering matrix formalism. 
The result 
is shown in Figure \ref{fig:correlation-surface}: The effect of the constraint \eqref{eq:momenta-corr-delta} is to give non-trivial correlations only on a surface in the $(k_y,k_y',k_x)$ space. The fact that this surface folds above itself for higher transverse momenta is due to the superluminal behaviour of the Bogoliubov dispersion, which imposes the upper bound \eqref{eq:dispersive-threshold} on the transverse momenta $k_x$ for which superradiant scattering and, hence, spontaneous emission can occur. 
The peculiar shape of this correlation surface provides an ideal signature to be looked after in experimental investigations.

From a quantitative point of view, the magnitude of the correlation function in \eqref{eq:momentumcorr-general} is largest at small $k_x$ where the emission intensity is strongest. However, as it typically occurs in two-mode parametric emitters~\cite{QuantumOptics}, the normalized correlation function
\begin{equation}
\label{eq:g2norm}
	g^{(2)}(k_y,k_y')=\frac{G^{(2)}(k_y,k_y')}{G^{(1)}(k_y,k_y)G^{(1)}(k_y',k_y')},
\end{equation}
that is considered in the colorplot, gives a maximum signal on the highest available values of $k_x$ for which the emission is the weakest in intensity but maintain equally strong quantum correlations. As compared to the three-particle nature of the Hawking emission process~\cite{busch2014}, the fact that the superradiant emission involves a single pair of opposite-norm modes is a favourable feature in view of detecting quantum entanglement between the emitted phonons~\cite{finazzi2014entangled}.

\section{Conclusions}
\label{sec:conclu}

In this article we have reported a theoretical study of the quantum superradiant emission from ergosurfaces in curved spacetimes. Such a phenomenon had originally been predicted for rotating black holes in a gravitational context but so far has escaped experimental verification. Here we investigate this effect in the context of analog models of gravity based on Bose-Einstein condensates.

In contrast to the draining bathtub vortex geometry considered in most earlier works on this subject, we focus here on the configuration originally proposed in~\cite{giacomelli2021understanding}, where a rotational velocity field is obtained in a planar geometry by applying a synthetic vector potential to the atomic gas. This is a minimal toy model that allows to disentangle superradiant scattering from other effects such as superradiant instabilities and black hole horizons and focus on its basic physics.

Our theoretical framework is based on a high-dimensional generalization of the input-output techniques originally developed in \cite{recati2009bogoliubov} for the study of analog Hawking radiation in one-dimensional condensates. This approach is based on the scattering matrix connecting the amplitudes of outgoing waves to the ones of the ingoing waves, which is directly obtained from the scattering solutions of the classical wave equation. Spontaneous emission of pairs of phonons then naturally arises as the superradiant scattering of the zero-point quantum fluctuations in the ingoing modes and its spectral distribution can be extracted from the reflection and transmission coefficients relating the different ingoing and outgoing modes.

A crucial advantage of our proposal is the simplicity of the geometry that, differently from rotating configurations, allows to efficiently track both components of the spontaneously emitted phonon pair. This was a key feature of the recent experimental reports of analog Hawking emission, whose main signature consisted in specific correlation features between opposite sides of the horizon~\cite{steinhauer2016observation,de2019observation}. 
Here, different kinds of correlation functions are considered: While the signature of superradiant emission in the position-space  density correlations may be geometrically too complex to be of effective use in experiments, an interesting and easily recognizable pattern is anticipated for the momentum-space two-particle correlation, another quantity of direct experimental access.

Even though our discussion focused on most celebrated case of analog models based on atomic Bose--Einstein condensates, our discussion can be straightforwardly applied also to analog models based on quantum fluids of light \cite{carusotto2013quantum}, in which both analog gravity~\cite{Gerace_2012,nguyen2015,jacquet2020polariton} and synthetic gauge fields~\cite{lim2017electrically,ozawa2019topological} are under very active consideration.

From a broader perspective of quantum field theories, the interest of our study goes well beyond superradiance. As it was discussed in~\cite{giacomelli2021understanding}, the planar quantum field configuration considered in our work can be exactly mapped onto a charged massive scalar field in one dimension coupled to an electrostatic potential. Our spontaneous emission processes can then be seen as an analog model of the spontaneous quantum emission associated to the bosonic Klein paradox \cite{nikishov1970barrier}. As such, they can be qualitatively related to the superradiance of charged non-rotating black holes~\cite{brito2020superradiance} and their associated quantum emission~\cite{gibbons1975vacuum}. The correlation features anticipated in our work then constitute a promising way to shine light on this broad class of phenomena of astrophysical and gravitational interest.

From a more speculative perspective, the advances reported in this work lay the ground for further investigations of more subtle superradiant effects. These include studies of quantum aspects of superradiant instabilities~\cite{kang1997quantum} and of the interplay of superradiant emission with Hawking radiation when both an ergosurface and a horizon are present, as well as more speculative explorations of the non-trivial interactions between spontaneous quantum emission, the quantum fluctuations of quasinormal modes, and the backreaction of the quantum fluctuations on the underlying background spacetime. As we have shown in this work, the flexibility of analog models based on quantum fluids and the additional possibilities offered by synthetic gauge fields are very promising assets in view of using analog model experiments to conceptually unravel the different physical effects that are at play in complex configurations of gravitational and cosmological interest.

\section*{Acknowledgements}
This work received funding from the European Union Horizon 2020 research and innovation program under Grant Agreement No. 820392 (PhoQuS) and from the Provincia Autonoma
di Trento.

\bibliography{biblio.bib}

\end{document}